\documentstyle[psfig,12pt]{article}
\def\ie{{\it i.e.}}
\def\lya{Lyman~$\alpha$}
\def\cm2{cm$^{-2}$}
\def\dsh{D/H}

\def\dshpres{(D/H)$_{\rm{pre}\odot}$}
\def\dshqso{(D/H)$_{\rm{QSO}}$}
\def\dshism{(D/H)$_{\rm{ISM}}$}
\def\kms{${\rm km\,s}^{-1}$}
\def\eg{{\it e.g.}}
\def\hi{H$\,\sc i$}
\def\di{D$\,\sc i$}
\def\ni{N$\,\sc i$}
\def\oi{O$\,\sc i$}
\def\cii{C$\,\sc ii$}

\def\siii{Si$\,\sc ii$}

\def\feii{Fe$\,\sc ii$}
\def\mgii{Mg$\,\sc ii$}
\def\bib {\par\noindent \hangindent=20pt \hangafter=1}
\def\ga{\mathrel{\mathchoice {\vcenter{\offinterlineskip\halign{\hfil
$\displaystyle##$\hfil\cr>\cr\sim\cr}}}
{\vcenter{\offinterlineskip\halign{\hfil$\textstyle##$\hfil\cr>\cr\sim\cr}}}
{\vcenter{\offinterlineskip\halign{\hfil$\scriptstyle##$\hfil\cr>\cr\sim\cr}}}
{\vcenter{\offinterlineskip\halign{\hfil$\scriptscriptstyle##$\hfil
\cr>\cr\sim\cr}}}}}
\def\la{\mathrel{\mathchoice {\vcenter{\offinterlineskip\halign{\hfil
$\displaystyle##$\hfil\cr<\cr\sim\cr}}}
{\vcenter{\offinterlineskip\halign{\hfil$\textstyle##$\hfil\cr<\cr\sim\cr}}}
{\vcenter{\offinterlineskip\halign{\hfil$\scriptstyle##$\hfil\cr<\cr\sim\cr}}}
{\vcenter{\offinterlineskip\halign{\hfil$\scriptscriptstyle##$\hfil
\cr<\cr\sim\cr}}}}}

\begin{document} 



\begin{center} 
\begin{Large} 
{\bf Deuterium abundances}
\end{Large} 
\end{center}

\begin{center}

Martin~Lemoine~$^1$, 
Jean~Audouze~$^2$, 
Lotfi~Ben~Jaffel~$^2$, 
Paul~Feldman$^3$,
Roger~Ferlet~$^2$, 
Guillaume~H\'ebrard~$^{3,2}$, 
Edward~B.~Jenkins~$^4$, 
Christoforos~Mallouris~$^5$, 
Warren~Moos~$^3$, 
Kenneth~Sembach~$^3$, 
George~Sonneborn~$^6$, 
Alfred~Vidal-Madjar~$^2$, 
and~Donald~G.~York~$^5$

\end{center}

\ 

\begin{small}

1: DARC, UPR--176 CNRS, Observatoire de Paris-Meudon, F-92195 Meudon
C\'edex, France

2: Institut d'Astrophysique de Paris, CNRS, 98 bis boulevard Arago, 
   F-75014 Paris, France

3: Johns Hopkins University, Department of Physics and Astronomy, 
   3400 North Charles Street, Baltimore,  MD 21218, USA

4: Princeton University Observatory, Princeton, NJ 08544, USA

5: University of Chicago, Department of Astronomy and Astrophysics, 
   5640 South Ellis Avenue, Chicago, IL 60637, USA

6: NASA/Goddard Space Flight Center, Code 681, Greenbelt, MD 20771, USA

\end{small}

\vskip 1.5cm

\noindent
{\bf Abstract.}
We discuss the measurements of deuterium abundances in 
high redshift quasar absorbers, in the solar system and in the interstellar 
medium. We present new results that indicate spatial variations of the
deuterium abundance in the interstellar medium at the level of $\sim50$\%
over scales possibly as small as $\sim10$~pc, and discuss plausible causes
for the origin of these variations.
\vskip 1cm

\bigskip


\newpage

\section{Introduction}

  In the early days of Big-Bang nucleosynthesis (BBN), starting with Alpher,
Bethe \& Gamov~(1948), and until the late sixties, the primordial origin of
$^4$He seemed quite plausible, but the site of formation of the other light
elements remained slightly mysterious. Reeves, Audouze,
Fowler \& Schramm (1973) argued for cosmological deuterium, and showed that a
baryonic density $\Omega_{\rm b}=0.016\pm0.005h^{-2}$ (with
$H_0=100h\,$km/s/Mpc) could explain the primordial abundance of $^2$D,
$^3$He, $^4$He, and possibly some $^7$Li. And, following calculations by
Truran \& Cameron (1971), they argued that in the absence of post-Big-Bang
production, deuterium is slowly destroyed during galactic evolution, as it is
entirely burned to $^3$He in stars; in particular, Truran \& Cameron (1971)
estimated a destruction factor $\sim2$. Reeves, Audouze, Fowler \& Schramm
(1973) thus argued that deuterium, if
solely produced in the Big-Bang, would be a monitor of stellar formation.
These ideas have been strengthened in the past twenty five years, and hardly,
if at all, modified: they form the current picture of the cosmological
significance of deuterium and its cosmic evolution.

  Notably Epstein, Lattimer \& Schramm~(1976) showed that no deuterium should
be produced in significant quantities in astrophysical sites other than the
Big-Bang. Hence, measured abundances of deuterium 
would provide lower limits to the primordial abundance and consequently, an
upper limit to the cosmic baryon density. It has been long recognized
that the primordial abundance of deuterium represents the most sensitive
probe of the baryonic density $\Omega_{\rm b}$ (see \eg,
Schramm, 1998; Schramm \& Turner~1998).

Until the late sixties, deuterium had only been detected in ocean water, at a
level \dsh$\sim10^{-4}$. In the early seventies, 
Black~(1971) and Geiss \& Reeves~(1972) 
performed the first indirect measurement of the abundance of
deuterium representative of the presolar nebula using 
combined solar wind and meteorite $^3$He measurements. 
Shortly after, Cesarsky et al.~(1973) attempted a detection via
the radio observation of the 21~cm and 92~cm lines of both \hi\ and
\di, and Rogerson \& York~(1973) successfully 
measured for the first time the abundance 
of deuterium in the interstellar medium from  
\hi\ and \di\ Lyman absorption lines. These  efforts were followed 
by numerous new studies over the following 25~years. 

During 
the past several years, measurements of the deuterium to hydrogen ratio in
moderate to high redshift absorbers toward quasars have been obtained for the 
first time. These clouds
are very metal-deficient, so that their deuterium  content should not have been
affected by astration of gas or, equivalently, the deuterium abundance
measured should be close to primordial. 
This is in contrast with the presolar nebula and
interstellar medium measurements, whose deuterium abundances show the
imprint of chemical evolution on the primordial abundance.

Therefore, we now have at our disposal three samples of 
deuterium abundances (measured by number in comparison with hydrogen),
each representative of a given epoch: BBN [primordial abundance
\dshqso], 4.5~Gyrs past [pre-solar abundance \dshpres] and present
epoch [interstellar abundance \dshism]. Note that the inference of a
primordial, pre-solar or interstellar  \dsh\ ratio from a measurement
rests on the assumption of efficient mixing of the material probed by
the observations.

Ultimately, we would like both to know the primordial \dsh\ 
ratio and to understand the evolution of its abundance with time, in order to 
constrain
the overall amount of star formation. As we discuss here, we have not yet
reached this goal. Interstellar measurements do not always agree with each
other and we argue, on the basis of very recent data, that at least 
part of the scatter is real; in other words, we argue that there exist some 
unknown processes that affect the \dsh\ ratio in the ISM by
$\sim30-50$\% 
in some cases, over possibly very small scales, and we discuss a
few plausible causes. In the case of the presolar nebula abundances, there also
exists scatter, but at the present time, it is not clear whether it 
arises from
chemical fractionation of deuterium and hydrogen in molecules, or from some
other cause. For quasar absorption systems, the situation is not yet clear,
although two remarkable measurements of Burles \& Tytler (1998a,b,c,d)
agree to a
common value \dshqso$=3.4\pm0.3\times10^{-5}$. As we have learned in the case
of the ISM, the picture may very well change when new observations come in
and we prefer to remain very cautious here.

In this paper, we discuss briefly the current determinations of deuterium
abundances and focus on the latest results from interstellar measurements. We
discuss QSO absorbers in Section~2, presolar nebula measurements in Section~3 
and ISM measurements in Section~4. Section~5 discusses possible causes of
spatial variations of the \dshism\ ratio and Section~6, their consequences on 
other estimates of the deuterium abundance. 
Finally Section~7 summarizes the conclusions and discusses future directions. 

\section{Primordial abundance}

Measurements of the \dsh\ ratio in metal-deficient absorbers on lines
of sight to distant quasars offer direct access to the primordial
abundance of deuterium (Adams~1976). Although of fundamental importance
with respect to Big-Bang nucleosynthesis, this measurement is
particularly difficult to achieve (Webb~et al.~1991). In the Lyman
series of ground state absorption by atomic \hi\ and \di\, the
absorption of deuterium appears 82~\kms bluewards (shorter wavelengths)
of the corresponding \hi\ absorption. In realistic situations, there is
only a limited range of $b$-values (the physical parameter that roughly
defines the width of the absorption line and which is related to the
temperature and the turbulent velocity) and column densities, for which
the absorption due to \di\ can be well separated from that of \hi.
Typically, for a single absorber and \hi\ column densities
N(\hi)$\sim10^{18}\,$\cm2, one would like the \hi\ $b$-value to range
around $\sim15\,$\kms, corresponding to temperatures $\sim10^4\,$K
(Webb et al.~1991; Jenkins~1996).  Such $b$-values are typical of
diffuse ISM clouds, but quite atypical of quasar absorbers, in which
the broadening parameter takes values above $\simeq15-20\,$\kms.
Moreover, one rarely observes a single absorber. In particular, the
Lyman~$\alpha$ forest is present at high redshifts $z\geq2$, with a
large density of lines per unit redshift, so that one has to
disentangle the \hi\ and \di\ from the numerous neighbouring weak lines
of \hi. In particular, one always runs the risk of confusion between a
\di\ line and a weak \hi\ line, at a redshift such that the line falls
at the expected position of the \di\ line; such \hi\ lines are called
interlopers. As a consequence, measuring the \dshqso\ ratio is a matter
of statistics.  Burles \& Tytler~(1998c,d) have estimated that about
one out of thirty quasars could offer a suitable candidate for a
measurement of the \dsh\ ratio.

The first upper limit on \dshqso\ was actually obtained by
York et al.~(1983) toward Mrk509, $z_{\rm abs}=0.03$ using IUE data.
Several years ago, Carswell et al. (1994) and Songaila et al. (1994) 
reported detections of deuterium absorption toward QSO0014+813,
\dshqso$\simeq25\times10^{-5}$ at
 $z_{\rm abs}=3.32$,
using respectively the Kitt Peak and W.M. Keck telescopes. These authors
were cautious in pointing out the possibility that the deuterium feature 
could actually be due to an \hi\ interloper. A new analysis of the Keck
data gave however \dshqso$\sim19\pm5\times10^{-5}$ (Rugers \& Hogan~1996). 
In the subsequent years, the situation has rapidly evolved. 
Our aim here is not to review all of these developments, and we
refer the reader to excellent existing reviews 
(Hogan~1997; Burles \& Tytler~1998c,d).
As of today, there are three strong claims for a detection of \di, namely 
\dshqso$=3.3\pm0.3\times10^{-5}$ at $z_{\rm abs}=3.57$ toward QSO1937-1009 
(Burles \& Tytler~1998a), 
\dshqso$=4.0\pm0.7\times10^{-5}$ at $z_{\rm abs}=2.50$ toward QSO1009+2956 
(Burles \& Tytler~1998b), and
\dshqso$=25.\pm10.\times10^{-5}$ at $z_{\rm abs}=0.701$ toward QSO1718+4807
(Webb et al.~1997). We therefore have two low and one high values of the
\dshqso\ ratio.

Here we note a few important points. From new observations of
QSO0014+813, Burles, Kirkman \& Tytler~(1999)
have demonstrated the presence of an \hi\ interloper in the absorption line
that had been identified as \di, so that consequently, no \dshqso\ ratio
could be measured with confidence in this system. However, Hogan (1998) 
maintains that there is evidence for a high deuterium abundance in this
system and that the probability and amount of contamination should be
small, basing his arguments on statistical studies of correlations of
absorbers on scales $\sim80\,$\kms. Songaila~(1998) 
reports a similar finding,
from statistical arguments, although based on a relatively small number of
lines of sight, and derives \dshqso$\geq5\times10^{-5}$. She also claims
that the estimate of the \hi\ column density of Burles \& Tytler toward
QSO1937-1009 is incorrect and finds for this system
\dshqso$\geq5\times10^{-5}$ (see however, Burles \& Tytler~1998e). 

We also note that Tytler et al.~(1999) 
have reanalyzed the HST data of QSO1718+4807 together with IUE and Keck 
spectra and concluded that, for a single 
absorber, \dshqso$=8-57\times10^{-5}$. However, they find that if a 
second \hi\ absorber is allowed for on this line of sight,
then the \dshqso\ ratio becomes 
an upper limit, \dshqso$\leq50\times10^{-5}$. On the other hand, using
Monte-Carlo simulations of \hi\ cloud distribution on the line of sight, they
could check  that the low \dshqso\ ratios toward QSO1009+2956 and
QSO1937-1009 held. Therefore the result toward QSO1718+4807 is not yet
conclusive; in particular, the HST dataset contains only \lya\ and an
associated Si{\,\sc iii} line and it would be extremely valuable to have data
on the whole Lyman series of this absorber. 

Finally, we note that Levshakov~(1998, for a review) suggests that
correlations in turbulent velocity on large spatial scales could
seriously affect determinations of the \dshqso\ ratio. This author, and
collaborators, claim that the above high and low measurements of the
deuterium abundance are consistent with a single value
\dsh$\simeq3.5-5.2\times10^{-5}$ (see also Levshakov, Tytler \& Burles~1999).

This field is too young and dynamic to permit highly confident conclusions at 
this time, although a trend toward \dshqso$\sim3.5\times10^{-5}$ seems to be 
emerging as indicated by the recent results of 
Burles \& Tytler~(1998a,b). 
Finally we  stress the need for further measurements of the 
\dshqso\ ratio, as they could change our understanding of the situation. 
This should be clear from our forthcoming discussion of the measurements of 
the \dshism\ ratio.

\section{Pre-solar abundance}

By measuring the $^3$He abundance in the solar wind, Geiss~\& Reeves~(1972) 
determined the abundance in the protosolar nebula and hence found 
\dshpres$\simeq2.5\pm1.0\times10^{-5}$. 
This result was historically the first evaluation of the deuterium 
abundance of astrophysical significance. 
It was confirmed by Gautier \& Morel~(1997) who showed 
\dshpres$=3.01\pm0.17\times10^{-5}$. These determinations of \dshpres\ are 
indirect and linked to the solar ($^4$He/$^3$He) 
ratio and its evolution since the formation of the solar system.

Whereas in cometary water deuterium is enriched by a factor of at least
10 relative to the protosolar ratio (\eg\ Bockel\'ee-Morvan et al.~1998;
Meier et al.~1998), 
the giant planets Jupiter and Saturn are considered to be undisturbed 
deuterium reservoirs, free from production or loss processes. 
Thus they should reflect the 
abundance of their light elements at the time of the formation of the solar system 
4.5~Gyrs ago (Owen et al.~1986). The first measurements of the \dshpres\ 
ratio in the Jovian atmosphere have been performed through methane and its 
deuterated counterpart CH$_3$D, yielding \dshpres$=5.1\pm2.2\times10^{-5}$ 
(Beer~\& Taylor~1973). Other molecules such as HD and H$_2$,  yield 
lower values: \dshpres$=1.-2.9\times10^{-5}$~(Smith et al.~1989). 

Recently, new measurements of the \dshpres\ ratio using very 
different methods were carried out. Two are based on the first results of the 
far infrared ISO observations of the HD molecule in Jupiter 
(Encrenaz et al.~1996) and Saturn (Griffin et al.~1996), and lead 
respectively to \dshpres$=2.2\pm0.5\times10^{-5}$ and 
\dshpres$=2.3^{+1.2}_{-0.8}\times10^{-5}$. 
Note that the Encrenaz et al.~(1996) value was updated to the more reliable 
value \dshpres$=1.8^{+1.1}_{-0.5}\times10^{-5}$ by Lellouch et al.~(1997). 
Another is based on the direct 
observation with HST-GHRS of both \hi\ and \di\ 
\lya\ emission at the limb of Jupiter for the first time (Ben~Jaffel 
et al.~1994;~1997) yielding \dshpres$=5.9\pm1.4\times10^{-5}$. 
The third one is an in situ measurement with a mass 
spectrometer onboard the Galileo probe (Niemann et al.~1996) yielding 
\dshpres$=5.0\pm2.0\times10^{-5}$ [however this last value has been revised 
recently toward the lower part of the range, \ie\ 
\dshpres$=2.7\pm0.6\times10^{-5}$ (Mahaffy et al.~1998)]. 

It is surprising that measurements that probe almost the same atmospheric 
region of Jupiter ($\sim1$~bar level) lead to a such a large scatter in 
the \dsh\ ratio. Indeed, the atmospheric composition at that level is the key 
parameter in the ISO data analysis, the H and D \lya\ spectra modeling and 
the Galileo mass spectrometer measurements.   

It is likely that the differences between these values are due to 
systematic effect associated with models, such as the CH$_4$ mixing 
ratio (Lecluse et al.~1996), the effect of aerosols, the effect 
of eddy diffusion, or in the case of the mass spectrometer data, 
instrumental uncertainties. Additional investigations and observations
including HST-STIS and FUSE observations will help to resolve this issue.

\section{Interstellar abundance}

The first measurement of the interstellar \dsh\ ratio was reported by
Rogerson \& York~(1973), from {\it Copernicus} observations of the line of
sight to $\beta$~Cen, giving \dshism$=1.4\pm0.2\times10^{-5}$. 
 In the subsequent
years, many other measurements of the interstellar deuterium abundances
were carried out from {\it Copernicus} and IUE observations of the
Lyman series of atomic \di\ and \hi\ (for a review, see \eg\ 
Vidal-Madjar, Ferlet \& Lemoine~1998). 
Because absorption by the Lyman series
takes place in the far-UV, these measurements require 
satellite-borne instruments, and the latest observations have been performed 
using HST and the {\it Interstellar Medium Absorption Profile Spectrograph}
(IMAPS), which afford higher spectral resolution.

In order to measure \dshism, one can also 
observe deuterated molecules such as HD, DCN, {\it etc}, and form the
ratio of the deuterated molecule column density to its non-deuterated
counterpart (H$_2$, HCN, {\it etc}). More than twenty different deuterated
species have been identified in the ISM, with abundances relative to the
non-deuterated counterpart ranging from $10^{-2}$ to $10^{-6}$.
Conversely, this means that fractionation effects are important. 
As a consequence, this method cannot currently provide a precise estimate of 
the true interstellar \dsh\ ratio. Rather, this method is used in conjunction
with estimates of the \dshism\ ratio to gather information on the
chemistry of the ISM. 

Another way to derive the \dshism\ ratio comes through radio observations
of the hyperfine line of \di\ at 92~cm. The detection of this line is
extremely difficult, but it would allow one to probe more distant interstellar
media than the local medium discussed below. However, because a large
column density of D is necessary to provide even a weak spin-flip
transition, these observations aim at molecular complexes. As a result,
the upper limit derived toward Cas~A (Heiles et al.~1993)
\dshism$\leq2.1\times10^{-6}$ may as well result from a large differential
fraction of D and H being in molecular form in these clouds, as from the
fact that one expects the \dsh\ ratio to be lower closer to the galactic
center (since D is destroyed by stellar processing). The most recent result 
is the low significance detection of interstellar \di\ 92~cm emission 
performed by Chengalur et al.~(1997) toward the galactic anticenter, 
giving \dshism$=3.9\pm1.0\times10^{-5}$. 

\ 

Therefore, the most reliable estimate of \dshism\ remains the
observation of the atomic transitions of D and H of the Lyman series in
the far-UV.  The relatively low resolution of the {\it Copernicus}
spectra ($\sim$15~\kms) usually left the velocity structure unresolved,
which could lead to significant errors. These uncertainties were
reduced when HST and IMAPS echelle observations provided resolving
powers high enough (3.5 to 4~\kms) to unveil the velocity structure.

Either the \lya\ lines emissions from cool stars or the continua from hot 
stars have been used as background sources.  Whereas cool stars can be
selected in the solar vicinity, luminous hot stars are
located further away, with distances 
$\ga100$~pc. Therefore, the line of sight to hot stars generally comprises
more absorbing components than cool stars. However, for
cool stars, the modeling of the stellar flux is usually much more
difficult than for hot stars. Moreover, lines of species such as
\ni\ and \oi\, that lie close to \lya\ cannot be observed, as the
flux drops to zero on either side of \lya. Hence, in the
case of cool stars, the line of sight velocity structure in \hi\ typically has 
to be traced with \feii\ and \mgii\ ions and this is usually not 
a good approximation. In contrast, \ni\ and \oi\ were shown to be good
tracers of \hi\ in the ISM (Ferlet~1981; York et al.~1983; 
Meyer, Cardelli \& Sofia~1997; Meyer, Jura \& Cardelli~1998;
Sofia \& Jenkins~1998) and hot stars are particularly interesting targets in that
respect. 

In any case, both types of background sources 
have offered some remarkable results. In the direction to the 
cool star Capella, Linsky~et al.~(1993; 1995) have obtained, using HST: 
\dshism$=1.60\pm0.09^{+0.05}_{-0.10}\times10^{-5}$. On this line of sight, 
only one absorbing component was detected, the Local Interstellar Cloud 
(LIC), in which the solar system is embedded (Lallement \& Bertin~1992). 
Several more cool 
stars have been observed with HST, all compatible with the Capella evaluation 
(Linsky et al.~1995: Procyon; Linsky \& 
Wood~1996: $\alpha$~Cen~A, $\alpha$~Cen~B; Piskunov et al.~1997: HR~1099, 
31~Com, $\beta$~Cet, $\beta$~Cas; Dring et al.~1997: $\beta$~Cas, 
$\alpha$~Tri, $\epsilon$~Eri, $\sigma$~Gem, $\beta$~Gem, 31~Com). The most 
precise of these measurements has been obtained toward HR~1099 by 
Piskunov et al.~(1997): \dshism$=1.46\pm0.09\times10^{-5}$. 
None of the other results is accurate enough to place any new 
constraints on the Linsky~et al.~(1993; 1995) evaluation. 

\ 

Recently, new observations by HST and IMAPS have become available. 
HST observations of white dwarfs instead of hot or cool stars can be used to 
circumvent most of the afore-mentionned difficulties. White dwarfs can be 
chosen near the Sun and they can also be chosen in the high temperature
range, so as to provide a smooth stellar profile at \lya. At the same
time, the \ni\ triplet at 1200~\AA\ as well as the \oi\ line at 1302~\AA\
are available. Such observations have now been conducted using HST toward 
three white dwarfs: G191-B2B (Lemoine et al.~1996; Vidal-Madjar et al.~1998),
Hz~43 (Landsman et al.~1996) and Sirius~B (H\'ebrard et al.~1999). 

Toward G191-B2B, Vidal-Madjar et al.~(1998) detected three absorbing
clouds using HST-GHRS 3.5~\kms\ spectral resolution data. Assuming that
all three absorbing components shared the same \dshism\ ratio, they
measured at \lya\ \dshism$=1.12\pm0.08\times10^{-5}$. There is a clear
discrepancy between this ratio and that observed toward Capella by
Linsky~et al.~(1993; 1995). As it turns out, one of the three absorbers
seen toward G191-B2B is the LIC, also seen toward Capella. Moreover,
the angular separation of both targets is 7$^o$. One should thus expect
to see the same \dshism\ ratio in both LIC line of sights. When this
constraint is included in the three-component fit, Vidal-Madjar et
al.~(1998) find that the average \dshism\ in the other two absorbers is
$\sim0.9\pm0.1\times10^{-5}$. Finally, it is important to note that
Vidal-Madjar et al.~(1998) re-analyzed the dataset of Linsky~et
al.~(1993; 1995) toward Capella, using the same method of analysis as
toward G191-B2B and confirmed the previous estimate. Therefore, the
conclusion is that the \dshism\ ratio varies by at least $\sim30\%$
within the local interstellar medium, either from cloud to cloud,
and/or within the LIC.

Using HST-GHRS observations, H\'ebrard et al.~(1999) detected two
interstellar clouds toward Sirius~A and its white dwarf companion
Sirius~B, one of them being identified as the LIC, in agreement with
previous HST observation of Sirius~A by Lallement et al.~(1994). As in
the case of G191-B2B, the interstellar structure of this sightline,
which is assumed to be the same toward the two stars (separated by less
than 4~arcsec at the time of the observation), is constrained by high
spectral resolution data of species such as \oi, \ni, \siii\ or \cii.
Whereas the deuterium \lya\ line is well detected in the LIC with an
abundance in agreement with the one of Linsky et al.~(1993, 1995), no
significant \di\ line is detected in the other cloud. However, the
\lya\ lines toward Sirius~A and Sirius~B are not simple. Indeed an
excess of absorption is seen in the blue wing of the Sirius~A
\lya\ line and interpreted as the wind from Sirius~A. In its white
dwarf companion, an excess in absorption is seen in the red wing and
interpreted as the core of the Sirius~B photospheric \lya\ line. A
composite \lya\ profile could nonetheless be constructed and the
\dshism\ measured in the second cloud is
\dshism$=0.5^{+1.1}_{-0.5}\times10^{-5}$ (90\% confidence level). The
rather large error bar stems primarily from the fact that only medium
resolution data were available for the \lya\ region.

Finally, IMAPS on the space shuttle ORFEUS-SPAS~II mission was used by
Jenkins et al.~(1999) to observe at high spectral resolution (4~\kms)
the Lyman~$\delta$ and Lyman~$\epsilon$ lines toward $\delta$~Ori.
These data allowed an accurate measurement of the \di\ column density.
Together with a new and accurate measurement of the \hi\ column density
from \lya\ spectra of $\delta$~Ori in the IUE archive, Jenkins et
al.~(1999) found the value \dshism$=0.74^{+0.19}_{-0.13}\times10^{-5}$,
at a 90\% confidence level (c.l.), which confirms the {\it Copernicus}
result obtained by Laurent et al.~(1979).  Compared to Capella
(Linsky~et al.~1993; 1995) and HR~1099 (Piskunov et al.~1997), this
value is very low.  This suggests that variations by $\sim$50\% are
possible in the local interstellar medium.

Using the same analysis techniques as Jenkins et al.~(1999), 
IMAPS was also used combined with IUE archive toward two other stars, 
$\gamma^2$~Vel and $\zeta$~Pup to yield the first results 
\dshism$=2.1^{+0.36}_{-0.30}\times10^{-5}$ and 
\dshism$=1.6^{+0.28}_{-0.23}\times10^{-5}$, respectively (Jenkins et al.~1998; 
Sonneborn et al.~1999). The value for $\gamma^2$~Vel is marginally inconsistent
with the lower value toward Capella, and this disparity may be substantiated
further when the error estimates become more refined. We also note that the
$\gamma^2$~Vel result confirms previous estimates of York \& Rogerson~(1976), 
while the $\zeta$~Pup result is only in marginal agreement with the
Vidal-Madjar et al.~(1977) evaluation, both made with {\it Copernicus}.

\section{Interstellar \dsh\ variations}

As we have discussed, there is now firm evidence for variations of the 
\dshism\ ratio, able to reach $\sim50$\%, 
over scales as small as $\sim10$~pc. This fact 
had already been suggested by early {\it Copernicus} and IUE data, although 
it was not known whether this was due to the inadequacy of the data and the
complexity of the problem, or to real physical effects. The dispersion of  
all published \dshism\ ratios, ranging from $0.5\times10^{-5}$ to 
$4\times10^{-5}$, was thus not universally accepted as real (Mc Cullough~1992).
Even if
some of this scatter may be accounted for by systematic errors, as we have
argued above, we  believe that at least part of it is real.

Actually, one should recall that time variations of the \dshism\ ratio
have already been reported toward $\epsilon$~Per (Gry et al.~1983).
They were interpreted as the ejection of high velocity hydrogen atoms
from the star, which would contaminate the deuterium feature.  Such an
effect can only mimic an enhancement the \dsh\ ratio, and it is thus
worth noting that in at least five cases, the \dshism\ ratio was found
to be really low:  $0.9\pm0.1\times10^{-5}$ in two components toward
G191-B2B (Vidal-Madjar et al.~1998, see Section~4);
$0.8\pm0.2\times10^{-5}$ toward $\lambda$~Sco (York~1983);
$0.5\pm0.3\times10^{-5}$ toward $\theta$~Car (Allen et al.~1992);
$0.7\pm0.2\times10^{-5}$ and $0.65\pm0.3\times10^{-5}$ toward
$\delta$~and $\epsilon$~Ori (Laurent et al.~1979), recently confirmed
in the case of $\delta$~Ori by Jenkins et al.~(1999):
\dshism$=0.74^{+0.19}_{-0.13}\times10^{-5}$ (90\% c.l.).  Two other
lines of sight seem to give low values for the \dsh\ ratio, albeit with
larger error bars: $0.5^{+1.1}_{-0.5}\times10^{-5}$ (90\% c.l.) in one
of the two components toward Sirius (H\'ebrard et al.~1999, see
Section~4), $0.8^{+0.7}_{-0.4}\times10^{-5}$ toward BD+28~4211 (G\"olz
et al.~1998).  All the above authors discussed possible systematics but
concluded that none of the identified ones could explain such low
values of \dshism.

Let us now discuss different plausible causes of variations:

\begin{itemize}
\item Molecular fractionation effects, such as the selective incorporation
of D into HD, {\it vs.} H into H$_2$, could modify the atomic \di/\hi\
ratio (Watson~1973). However, the absorbers mentioned above are not
molecular, with typical H$_2$/\hi\ ratios $\leq10^{-4}$, so that this should
not be a strong effect.

\item Vidal-Madjar et al.~(1978), and Bruston et al.~(1981) have
suggested that the anisotropic flux in the solar neighborhood, combined
with a differential effect of radiation pressure on \hi\ and
\di\ atoms, would result in the spatial segregation of \di\ {\it vs.}
\hi. Indeed, all \di\ atoms in a cloud with
N(\hi)$\sim10^{18}\,$\cm2\ are subject to resonant radiation pressure,
whereas the inner \hi\ atoms are shielded from the flux by the
optically thick \hi\ enveloppe. Therefore, provided that the cloud is
not homogeneous (and radiation is anisotropic), the segregation of
\di\ atoms {\it vs.} \hi\ atoms induces a spatial variation in the
\dsh\ ratio. In particular, they predict that the \dsh\ ratio would
appear either higher or lower than its actual value, depending on where
the line of sight crosses the cloud, assuming it is perpendicular to
the direction of the net radiation flux. There is however more chance
to observe the depleted region, since it is much more extended than the
enriched region. These authors calculate that for a flux corresponding
to 10 OB stars located at $50\,$pc from the cloud, deuterium atoms
could diffuse to the other side in a timescale $\sim10^6\,$yrs. The
clear signature of this mechanism would be the evidence of regional
differences in the ISM.

\item Jura (1982) has suggested that the adsorption of \di\ and \hi\ onto
dust grains could be selective. In this respect, it would be interesting to
study the variation of \dshism\ with gas velocity, as if there was indeed
such a correlation, in much the same manner as Ca$\,\sc ii$/Na$\,\sc i$ varies
(the Routly-Spitzer effect, Routly \& Spitzer~1952, Vallerga et al.~1993 and
references), one might be more willing to accept the conjecture
about the difference in binding of D and H to the surfaces of dust grains.
We note, however, that the measurement of N(\hi) cannot be done precisely in
absorption spectra on a component by component basis, if there are more than
one absorber, so that the velocity information in the \dsh\ ratio is lost. One
usually measures average \dsh\ ratios, such as for G191-B2B. Therefore, one
should measure D/O or D/N ratios as a function of velocity.

\item Copi, Schramm \& Turner (1995), and Copi (1997)
have devised a stochastic approach to chemical evolution, in which they
compute the evolution of a particular region of space in Monte-Carlo
fashion. This allows them to probe the scatter around the mean of
correlations such as abundances {\it vs.} time/metallicity; this is in contrast
to usual models of chemical evolution that only compute the mean behavior.
Actually, one of their objectives was to study the spread in light elements
abundances after 15~Gyrs of evolution, and they find that for deuterium, one
expects a negligible scatter. However, it is difficult to apply their results
to spatial variations of the \dshism\ ratio, as they were more concerned
with variations at given metallicities, and thus did not introduce spatial
dependence in their Monte-Carlo calculations.

\item Along similar lines of thought, let us see if stellar ejecta
could introduce inhomogeneities in the \dshism\ ratio. One should focus
on planetary nebula (PN) ejecta and cool giant winds, as their mass
input in the ISM dominates that of other stars (Pottasch~1983).
Moreover, PN ejecta and cool giant winds share similar characteristics
with interstellar clouds in which deuterium has been seen: PN ejecta
have mass $\sim10^{-2}-10^{-1}\,M_{\odot}$, and speed $\sim20\,$\kms;
cool giants have mass loss $\sim3\times10^{-6}\,M_{\odot}{\rm
yr}^{-1}$, with velocities $\sim10\,$\kms; above interstellar clouds
have mass $M\sim10^{-2}\,M_{\odot}$ (if $n_{\rm HI}\sim0.1\,$cm$^{-3}$
and N(\hi)$\sim10^{18}\,$\cm2), and speed $\sim10-20\,$\kms.  Therefore
the admixture of PN ejecta or giant wind, that are deuterium free (all
D is burned to $^3$He by pre-main sequence), and interstellar
unprocessed material, would result in a \dsh\ ratio reduced by $M_{\rm
cloud}/(M_{\rm poll}+M_{\rm cloud})$, where $M_{\rm poll}$ and $M_{\rm
cloud}$ denote the polluted and interstellar mass, respectively.

  The probability that a given line of sight crosses a PN ejecta is
given by the covering factor of PNe on a sphere of radius $R$, centered
on the Sun. The observed density of planetary nebulae in the solar
vicinity is $N_0\sim5\times10^{-8}\,$pc$^{-3}$; this however, counts
only visible nebulae, whose age is $\leq3\times10^4\,$yr.  Assuming
that the density of  nebulae of age $t$ scales as:  $N=N_0t/t_0$, where
$t_0$ is the age before disappearance, one obtains the covering factor
for nebulae of a given age $t$:  $f\approx
7\times10^{-4}N_{0,\,-7.3}R_{50}r_{10}^2(t/t_0)$, assuming $r\ll R$,
where $r_{10}$ is the radius of the PN in units of $10\,$pc, $R_{50}$
is in units of $50\,$pc, $N_{0,\,-7.3}$ in units of
$5\times10^{-8}\,$pc$^{-3}$; $r$ is tied to the age $t$, for instance
$r=20\,{\rm pc}\,v_{20}t_{6}$ ($t_{6}$ in units of $10^6\,$yrs), if the
expansion is linear. Since the covering factor grows as $r^2t$, one
only needs to consider the largest (oldest) PNe.  A rough estimate of
the maximum radius of expansion can be obtained by equating the dynamic
pressure of the ejecta and the ISM pressure:  $r\sim3\,{\rm
pc}\,n_{-1}^{-1/3}T_4^{1/3}M_{-2}^{1/3}v_{20}^{2/3}$, where $n_{-1}$ is
the total density of the ambient ISM in units of $0.1\,$cm$^{-3}$,
$T_4$ is the temperature in $10^4\,$K, $M_{-2}$ the ejecta mass in
$10^{-2}\,M_{\odot}$. This corresponds to a covering factor
$f\sim2\times10^{-4}$, for an age $t\sim10^5\,$yrs ({\it i.e.} assuming
linear expansion). Note that $r\sim3\,$pc roughly corresponds to the
typical size of an ISM cloud.

  One could reach higher covering factors in low density media, such as
the solar vicinity, where $n\sim10^{-4}\,$cm$^{-3}$ and $T\sim10^6\,$K
(Cox \& Reynolds~1987; Ferlet~1999): in this case, the maximum radius
of expansion of ejecta can become larger. However, if this latter
becomes much larger than the typical size of an interstellar cloud,
then the polluted mass that is effectively mixed with the interstellar
material is less than the ejected mass, and the mixing becomes
ineffective. Therefore, we feel that the above covering factor,
$f\sim2\times10^{-4}$, for $R=50\,$pc and $r\simeq3\,$pc, should give a
reasonable estimate of the probability of contamination, within an
order of magnitude.

  Finally, one can perform a similar calculation for cool giant winds.
Their number density is $\sim2.5\times10^{-7}\,$pc$^{-3}$; each  ejects
$\sim0.3\,M_{\odot}$ on a dynamical evolution timescale $\la10^6\,$yrs.
This corresponds to a covering factor $f\sim4\times10^{-3}$, which is
substantially larger than for PNe.  Although these are a qualitative
estimates, the probability of contamination appears marginal, but
cannot be ruled out either, and quite probably so for cool giant winds,
when long pathlengths are considered.

  The best signature of contamination of material of solar chemical
composition by a PN ejecta, or giant wind, comes through fluorine,
which usually shows [F/O]$\sim1$ (Kaler~1982 for PNe; Joriseen, Smith
\& Lambert~1992 for giants); elements such as C and N are not always
over-abundant.  Note, however, that fluorine may also be interpreted as
a signature of $\nu$-process in type II supernovae (Timmes et al.~1997,
and references). Fluorine may be detected in absorption with lines of
F{\sc i} at 952\AA\ and 954\AA, although its weak universal abundance
makes the detection rather difficult. Nonetheless, FUSE has access to
this range, and should thus offer a possibility to test such
contamination of interstellar material.

  \item  Even though it is always much easier to destroy deuterium than
to fabricate it in astrophysical systems, several processes that led to
production of deuterium have been mentioned. Note that  Epstein,
Lattimer \& Schramm (1976) showed that no realistic astrophysical
system could produce deuterium by {\it nucleosynthesis} or {\it
spallation} mechanisms, without, in the latter case, overproducing Li.
However, photodisintegration of $^4$He can lead to production of
deuterium, as exemplified by Boyd, Ferland \& Schramm (1989). These
authors showed that $\gamma$-ray sources associated with \eg, galactic
centers and/or AGNs, could photo-disintegrate $^4$He and lead to
significant production of, among others, deuterium. However, the radius
of influence of such processes is usually very small; in the above
case, it was found to be $\sim10\,$light years. This plus the rarity of
$\gamma$-ray sources in the Galaxy, makes the contamination of
interstellar abundances unlikely.

Jedamzik \& Fuller (1997) have studied the possibility of
photo-disintegrating $^4$He with high-redshift $\gamma$-ray bursts, and
conclude that this is highly improbable due to the small radius of
influence $\sim10\,$pc, a result similar to that of Boyd, Ferland \&
Schramm (1989).  Nonetheless, Cass\'e \& Vangioni-Flam~(1998; 1999)
have argued that blazars could actually influence absorbers in a
significant way if the absorber is a blob of matter expelled by the
central engine. Interestingly, they predict as a generic signature of
photo-disintegration that odd to even atomic number element ratios
should be super-solar, notably the N/O ratio.  They also argue that
creation as well as destruction of deuterium can occur, depending on
the $\gamma$-ray spectrum.

  \item Jedamzik \& Fuller~(1995, 1997) have pointed out that primordial
isocurvature baryon fluctuations on mass scales $\leq10^5-10^6\,M_{\odot}$
could produce variations by a factor 10 on these scales, and variations of
order unity on galactic mass scales $\sim10^{10}-10^{12}\,M_{\odot}$. However,
this attractive and original scenario would not apply to \dshism\ ratios on
very small spatial scales $\sim10\,$pc.

  \item Finally, Mullan \& Linsky~(1999) have argued that production of
deuterium in stellar flares, by radiative capture of a proton by a free
neutron, can produce a non-negligible source of non-primordial
deuterium in the ISM, and possibly explain the observed variations.
However, detailed estimates are still lacking for this scenario.

\end{itemize}

  Finally, one should note that the above mechanisms do not agree as to
whether D should be enhanced or depleted with respect to H, if any of
them operates. Therefore, one cannot conclude which one of the observed
interstellar abundances, if any, is more representative of a cosmic abundance
that would result solely from Big-Bang production followed by star formation.

\section{Discussion} 

  The next question that comes to mind is the following: taking for granted
that variations in the \dshism\ ratio exist, do we expect to see a similar
effect in QSO absorption line systems, and if yes, how would it affect the
estimate of the primordial abundance of deuterium?

  To be brief, we do not know the answer to this question, mainly because
the nature and the physical environment of absorption systems at high
redshift may be very different from that of interstellar clouds in the
solar neighborhood, and because we do not know the origin of the variation
of the \dshism\ ratio.

  The \dshism\ ratio is measured in interstellar clouds that typically show:
N(\hi)$\sim10^{18}\,$\cm2, $n_{\rm H}\sim0.1\,$cm$^{-3}$, ionization
$n_{\rm HI}\sim n_{\rm HII}$, $T\sim10^4\,$K, size $L\sim1\,$pc, and mass
$M\sim10^{-2}\,M_{\odot}$. Although
the Lyman limit systems have a similar column densities, their 
physical characteristics may be very different. One opinion is that
these systems are associated with extended gaseous haloes, as one
often finds a galaxy at the redshift of the absorber with an impact
parameter $R\sim30h^{-1}\,$kpc (Bergeron \& Boiss\'e~1991; Steidel~1993).
However, it is not known whether this absorption is continuous and extends on
scales $\sim R$, or whether the absorption is due to discrete clouds
sufficiently clustered on the scale $\sim R$ to produce absorption with
probability $\approx1$. In particular, York et al.~(1986) and Yanny \&
York (1992) have suggested that QSO Lyman~$\alpha$ absorption (not
necessarily Lyman limit systems) occur in clustered dwarf galaxies undergoing
merging. In this case, one expects the absorbers to be much like galactic
clouds, in particular, of small spatial extent. It is also usually believed
that the QSO UV background is responsible for the ionization properties of
these absorbers, in which case one typically derives low densitites
$n_H\sim10^{-3}\,$cm$^{-3}$, which translates into large masses
$\sim10^8\,M_{\odot}$ (plus or minus a few orders of magnitude) if the clouds
have a large radius $\sim30\,$kpc. However, there are other models for the
ionization of these Lyman limit systems. For instance, Viegas \&
Fria\c{c}a~(1995) have proposed a model where Lyman limit systems originate
in galactic haloes, have sizes $\sim {\rm few}\,$kpc, hydrogen densities
$n_H\sim10^{-1}$, and the ionization results from the surrounding hot gas. 
Lyman limit systems are shrouded in mystery.

  Despite these large uncertainties, one can establish a few
interesting points. First,
the depletion of deuterium by contamination of low mass stellar ejecta has
been ruled out by Jedamzik \& Fuller (1997). Indeed, the QSO absorbers where
D has been detected have been shown to be very metal-poor.
The metallicity inferred, typically [C/H]$\leq-2.0$, implies that no more than
1\% of the gas  been cycled through stars. We note that Timmes et al.~(1997)
suggest that the incomplete mixing and the smallness of the QSO beam could
introduce non-negligible variations in \dshqso\ ratios.

  Differential radiation pressure could affect
measured \dshqso\ ratios if Lyman limit systems are discrete clouds, and
their radius is not too large. Indeed, the primary requirement
of the model of Vidal-Madjar et al.~(1978) and Bruston et
al.~(1981), is that the radiation flux be anisotropic, and for maximum
efficiency, that the line of sight cross the absorber perpendicularly to
the direction of the radiation flux. As it turns out, QSO absorbers chosen
for measurements of \dsh\ ratios fulfil these criteria. In effect, these
sytems are selected for \dsh\ studies if their line of sight
is as trivial as possible. This means that the absorber has to be isolated,
which, in geometrical terms means, for a spherical distribution,
that it has to lie on the
boundary. If the radiation flux arises from the central part of the spherical
system, it is anisotropic on a cloud at the boundary,
and moreover, the line of sight is
effectively perpendicular to the flux impinging on the cloud; otherwise, one
would expect  multiple absorbers on the line of sight. Following Bruston 
et al.~(1981), the diffusion velocity of deuterium atoms is:
$v_{\rm D}\sim 1\,{\rm pc.Myr}^{-1} \Phi_{-6}n_{-1}^{-1}T_{4}^{-1/2}$, for
$\Phi_{-6}$ in units of $10^{-6}\,$photons/cm$^2$/s/Hz, $n_{-1}$ total density
in units of $0.1\,$atoms.cm$^{-3}$, and $T_{4}$ in units of $10^4\,$K. 

  In the case of the York {\it et al.}~(1986) model, one expects a
radiation flux corresponding to $\sim10^2-10^4$ O stars, impinging on a
cloud located $\sim1$kpc from the center of the dwarf galaxy. This
gives a diffusion velocity $v_{\rm D}\sim0.02-2\,$pc/Myr. Diffusion,
hence segregation, can thus occur over scales of a $\sim1\,$pc, a typical
cloud size, as the relevant timescale is the crossing time $\sim20\,$Myr
for a cloud circulation velocity $\sim50\,$km/s. However, for the model
developed by Viegas \& Fria\c{c}a~(1995), the typical diffusion
distance for deuterium atoms is $\sim1-10\,$pc, for a flux
$\Phi_{-6}\sim0.01-0.1$, sustained on a star formation timescale
$\sim10^8\,$yrs.  This distance is small compared to the modeled cloud
size $\sim{\rm few}\,$kpc, and one thus does not expect segregation of
deuterium.

  Although the above numbers are very qualitative, mainly because of
the uncertainties inherent to our knowledge of Lyman limit systems, one
cannot rule out an effect of anisotropic radiation. It is actually
interesting that the criteria according to which Lyman limit systems
are chosen for \dsh\ studies, coincide with those for a maximum effect
of radiation pressure. In moderately ionized and small sized
($\sim1-10\,$pc) regions, the deuterium abundance could be depleted by a
factor 2.

  Finally, we note that, whatever the right value of the primordial deuterium
abundance, that is, either low $\sim3.5\times10^{-5}$, or high
$\sim10^{-4}$, there is satisfying agreement with both Big-Bang
nucleosynthesis and the predictions of other light elements abundances, and
with chemical evolution and the interstellar abundances of deuterium.
See, {\it e.g.}, Schramm \& Turner (1998) for a discussion of the agreement
of a low \dshqso\ with primordial $^4$He determinations (and statistical
errors), and primordial $^7$Li, and its cosmological implications. High
deuterium abundances are known to provide very good agreement with BBN
predictions for $^4$He and $^7$Li. Although they predict significant
astration of deuterium: \dshqso/\dshism$\sim5-10$, it is also known there
exist viable chemical evolution models able to account for such a large
destruction ({\it e.g.} Vangioni-Flam \& Cass\'e~1995; Timmes et al.~(1997);
Scully et al.~1997).

\begin{figure}
\centerline{\psfig{figure=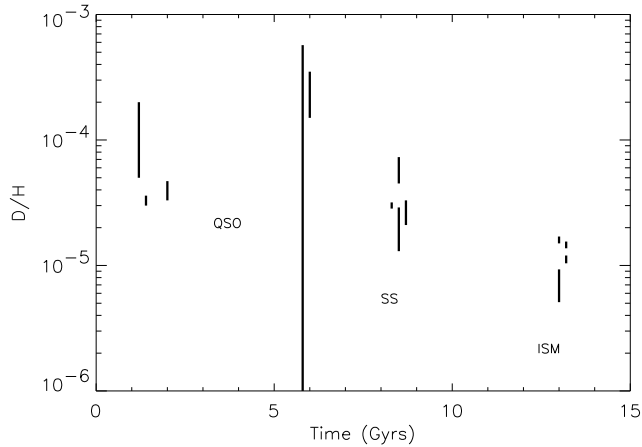,width=0.60\textwidth}}
\caption{{\bf Deuterium abundance measurements.} 
The different \dsh\ evaluations are shown as a function of time (for
$\Omega_0=1$, $q_0=0.5$, $H_0=50$ km/s/Mpc).
The primordial measurements plotted (QSO) are from
Burles \& Tytler~(1998a,b) 
and Songaila~(1997) [high redshifts] and from Webb et al.~(1997) 
and Tytler et al.~(1999) [moderate redshift]. Pre-solar values plotted 
(SS) are from Gautier \& Morel~(1997) [solar wind] and Lellouch et al.~(1997), 
Ben~Jaffel et al.~(1997) and Mahaffy et al.~(1998) [Jupiter]. 
Interstellar values 
(ISM) plotted are the ones from Linsky~et al.~(1995) [Capella], 
Piskunov et al.~(1997) [HR~1099], Vidal-Madar et al.~(1998) [G191-B2B] 
and Jenkins et al.~(1999) [$\delta$~Ori].}
\end{figure}

\section{Conclusion}

  The different \dsh\ evaluations reviewed or presented here are shown in
Fig.~1, as a function of (approximate) time. This figure seems to reveal a
trend of decreasing deuterium abundance with time, predicted as early as 1971
(Truran \& Cameron~1971). However, if one looks more closely at Fig.~1, there
are discrepancies between different evaluations of  deuterium abundances at
similar cosmic time, which, as we have argued in the case of ISM measurements,
cannot be always accounted for in terms of measurements systematics. 

  Nevertheless, the trend indicated in Fig.~1 seems to show that we
are converging toward a reasonable (at least understandable) picture of the
cosmic history of deuterium, or Deuteronomy, in Dave Schramm's own terms.

  We have the hope that FUSE, scheduled for launch in early 1999, will
sharpen this picture, and fill in the gaps to construct a curve of evolution
of the abundance of deuterium {\it vs} time/metallicity. The FUSE Science
Team intends to conduct a comprehensive study
of the deuterium abundance in the Galaxy through Lyman series 
absorption of \di\ between 912 and 1187~\AA.  Access to a suite
of lines in the series provides much stronger constraints on
$N$(\di) and $N$(\hi) than single line (\ie, \lya) observations
alone.  The bandpass also contains a large number of lines of \oi,
\ni, and \feii\ that can be used to trace the metallicity and dust
content of the absorbers studied.

The primary goal of the FUSE \dsh\ program is to link the destruction of 
deuterium to the physical and chemical properties of the interstellar gas.  
This objective is critical to successful galactic chemical evolution 
models since astration of deuterium, metal production, and mixing/recycling 
of the ISM are key ingredients in the models. FUSE observations of \dsh\ 
in environments with different chemical histories will help to 
reveal the effectiveness of astration and its dependence upon 
environmental factors (\eg\ metallicity, star-formation). A study of regional
variations may reveal evidence that supports the proposal about the
differential effect of radiation pressure. Finally, \dsh\ measurements
in regions of low metallicity will be particularly important benchmarks 
for relating the high redshift \dsh\ values to present epoch values.

FUSE will be capable of observing deuterium in distant gas clouds 
beyond the solar neighborhood clouds explored by {\it Copernicus}, HST, 
and IMAPS.  Therefore, it should be possible to search for large
scale variations in \dsh\ related to global star formation and 
metal gradients, as well as small scale variations in selected regions
due to incomplete mixing of the interstellar gas or deuterium decrements 
in the ejecta of stars.

FUSE, together with HST-STIS and IMAPS, 
should thus give access to more precise \dsh\ evaluations and 
should greatly clarify the problem of the chemical 
evolution of deuterium, and hence much better constrain our understanding
of the primordial \dsh\ value.

\

{\bf References :}

\bib   Adams, T. F.: 1976, A\&A 50, 461

\bib   Allen, M. M., Jenkins, E. B., \& Snow, T. P.: 1992, ApJS 83, 261 

\bib   Alpher, R. A., Bethe, H. A., \& Gamov, G.: 1948, Phys. Rev. 73, 803


\bib   Beer, R., \& Taylor, F.: 1973, ApJ 179, 309

\bib   Ben Jaffel, L., et al.: 1994, Bull. Am. Astron. Soc. 26, 1100

\bib   Ben Jaffel, L., et al.: 1997, {\it The Scientific return of GHRS}


\bib  Bergeron, J., Boiss\'e, P.: 1991, AA 243, 344

\bib   Black, D. C.: 1971, Nature Physic Sci. 234, 148

\bib   Bockel\'ee-Morvan, D., et al.: 1998, Icarus 133, 147


\bib  Boyd, R.N., Ferland, G.J., Schramm, D.N.: 1989, ApJ 336, L1

\bib    Bruston, P., Audouze, J., Vidal-Madjar, A., \& Laurent, C.: 1981, 
ApJ 243, 161

 \bib   Burbidge, G. \& Hoyle, F.: 1998, ApJ 509, in press

\bib Burles, S., Kirkman, D., Tytler, D.: 1999, ApJ 519, in press

 \bib   Burles, S., \& Tytler, D.: 1998a, ApJ 499, 699

 \bib   Burles, S., \& Tytler, D.: 1998b,  ApJ 507, 732

\bib    Burles, S., \& Tytler, D.: 1998c, , Proceedings of the 
Second Oak Ridge Symposium on Atomic \& Nuclear Astrophysics, Eds. A. 
Mezzacappa, {\tt astro-ph/9803071}

\bib  Burles, S., Tytler, D.: 1998d, Sp. Sc. Rev. 84, 65

\bib  Burles, S., Tytler, D.: 1998e, AJ 114, 1330

\bib    Carswell, R. F., Rauch, M., Weymann, R. J., Cooke, A. J., \& Webb, J. K.: 
1994, MNRAS 268, L1


 \bib   Cass\'e, M., \& Vangioni-Flam, E.: 1998, in Structure and Evolution of 
the Intergalactic Medium from QSO Absorption Line Systems, Eds. 
Petitjean, P., \& Charlot, S., 331

\bib  Cass\'e, M., Vangioni-Flam, E.: 1999, in $3^{rd}$ Integral Workshop
(ESA), Taormina Sept.1998, in press

\bib    Cesarsky, D.A., Moffet, A.T., \& Pasachoff, J.M.: 1973, ApJ 180, L1

\bib    Chengalur, J. N., Braun, R., \& Burton,  W. B.~: 1997, A\&A 318, L35

\bib    Copi, C. J., Schramm, D. N., \& Turner, M. S.: 1995, ApJ 455, L95 

\bib     Copi, C.J: 1997, ApJ 487, 704

\bib  Cox, D.P., Reynolds, R.J.: 1987, ARAA 25, 303


\bib    Dring, A.R., Linsky, J., Murthy, J., Henry, R. C., Moos, W., Vidal-Madjar, 
A., Audouze, J., \& Landsman, W.: 1997, ApJ 488, 760

\bib   Encrenaz, Th., et al.~: 1996, A\&A 315, L397

\bib   Epstein, R. I., Lattimer, J. M., \& Schramm, D.N.: 
1976, Nature 263, 198

\bib   Ferlet, R.: 1981, A\&A 98, L1


\bib  Ferlet, R.: 1999, ARAA, in press

\bib   Fuller, G. M., \& Shu, X.: 1997, ApJ 487, L25



\bib   Gautier, D., \& Morel, P.:1997, A\&A 323, L9

\bib   Geiss, J., \& Reeves, H.: 1972, A\&A 18, 126


\bib   G\"olz, M., et al.: 1998,  Proc. IAU Colloq. 166, The Local Bubble 
and Beyond, Eds. Breitschwerdt, D., Freyberg, M. J., Trumper, J., 75

\bib   Gnedin, N. Y., Ostriker, J. P.: 1992, ApJ 400, 1
 
\bib  Grevesse, N., Noels, A.: 1993, in Origin and evolution of the elements,
eds M. Cass\'e, N. Prantzos, E. Vangioni-Flam (CUP), p15

 \bib  Griffin, M. J., et al.~: 1996, A\&A 315, L389

 \bib  Gry, C., Laurent, C., \& Vidal-Madjar, A.: 1983, A\&A 124, 99

\bib   H\'ebrard, G., Mallouris, C., Vidal-Madjar, A., et al.: 1999, 
to be submitted to A\&A

\bib   Heiles, C., McCullough, P., \& Glassgold, A.: 1993, ApJS 89, 271

\bib   Hogan, C. J.: 1997, Proceedings of the ISSI workshop, Primordial
Nuclei and their Galactic Evolution, {\tt astro-ph/9712031}

\bib   Hogan, C. J.: 1998, Space Sci. Rev. 84, 127

\bib  Jedamzik, K., Fuller, G.M.: 1995, ApJ 452, 33

\bib  Jedamzik, K., Fuller, G.M.: 1997, ApJ 483, 560

\bib Jenkins, E.B.: 1996, in Cosmic Abundances, eds S.S. Holt, G. Sonneborn
(ASP Conf. Series, 99), p90

 \bib  Jenkins, E. B., et al.: 1998, in ESO workshop --
 Chemical Evolution from Zero to High Redshift, Oct. 14-16

\bib   Jenkins, E. B., Tripp, T. M., Wozniak, P. R., Sofia, U. J., \& 
Sonneborn, G.: 1999, submitted to ApJ, preprint astro-ph/9901403

\bib   Jura, M. A.: 1982, {\it Four Years of IUE Research}, NASA CP 2238, 54

\bib  Kingsburgh, R.L., Barlow, M.J.: 1994, MNRAS 271, 257

\bib   Lallement, R., \& Bertin, P.: 1992, A\&A 266, 479

\bib   Lallement, R., Bertin, P., Ferlet, R., Vidal-Madjar, A., \& Bertaux, J.L.: 
1994, A\&A 286, 898

\bib   Landsman, W., Sofia, U. J., \& Bergeron, P.: 1996, Science with 
the Hubble Space Telescope - II, STScI, 454

\bib   Laurent, C., Vidal-Madjar, A., \& York, D. G.: 1979, ApJ 229, 923

\bib   Lecluse, C., Robert, F., Gautier, D., \& Guiraud, M.: 1996, 
Plan. Space Sci. 44, 1579

\bib   Lellouch, E., et al.: 1997, Proceedings of the ISO workshop at Vispa, 
Oct. 97, ESA-SP 419, 131

\bib   Lemoine, M., Vidal-Madjar, A., Bertin, P., Ferlet, R., Gry, C., 
Lallement, R.: 1996, A\&A 308, 601

\bib  Levshakov, S.A.: 1998, to appear in the Proceedings of the Xth Rencontres 
de Blois, {\tt astro-ph/9808295}

\bib Levshakov, S.A., Tytler, D., Burles, S.: 1999, AJ submitted,
preprint astro-ph/9812114

\bib  Linsky, J. L., Brown, A., Gayley, K., Diplas, A., Savage, B. D., 
Ayres, T. R., Landsman, W., Shore, S. W., Heap, S. R.: 1993, ApJ 402, 694

\bib    Linsky, J. L., Diplas, A., Wood, B. E., Brown, A., Ayres, T. R., 
Savage, B. D.: 1995, ApJ 451, 335

 \bib   Linsky, J., \& Wood, B. E.: 1996, ApJ 463, 254

\bib McCullough, P.R.: 1992, ApJ 390, 213

\bib    Mahaffy, P. R., Donahue, T. M., Atreya, S. K., Owen, T. C., \& Nieman, 
H. B.: 1998, Space Sci. Rev. 84, 251

 \bib    Meier, R., et al.: 1998, Science 279, 842

\bib    Meyer, D.M., Cardelli, J.A., Sofia, U.J.: 1997, ApJ 490, L103

\bib Meyer, D.M., Jura, M., Cardelli, J.A.: 1998, ApJ 493, 222

\bib Mullan, D.J., Linsky, J.L.: 1999, ApJ 511, 502



 \bib    Niemann, H. B., et al.: 1996, Science 272, 846


\bib     Owen, T., Lutz, B., \& De Bergh, C.: 1986, Nature 320, 244

\bib     Piskunov, N., Wood, B. E., Linsky, J.L., Dempsey, R. C., \& Ayres, T. R.: 
1997, ApJ 474, 315

\bib    Pottasch, S.R.: 1983, Planetary Nebulae (D. Reidel, Astrophysics and
Space Science Library)



\bib     Reeves, H., Audouze, J., Fowler, W. A., \& Schramm, 
D. N.: 1973, ApJ 179, 909

\bib     Rogerson, J., \& York, D.: 1973, ApJ 186, L95

\bib Routly, P.M., Spitzer, L.: 1952, ApJ 115, 227

\bib     Rugers, M., Hogan, C.J.: 1996, ApJ 459, L1
 
\bib    Schramm, D.N.: 1998, Sp. Sc. Rev. 84, 3

\bib    Schramm, D.N., Turner, M.S.: 1998, Rev. Mod. Phys. 70, 303


 \bib    Scully, S. T., Cass\'e, M., Olive. K. A., \& 
Vangioni-Flam, E.: 1997, ApJ 476, 521

 \bib   Smith, WM. H., Schemp, W. V., \& Baines, K. H.: 1989, ApJ 336, 967

\bib Sofia U.J., Jenkins, E.B.: 1998, ApJ 499, 951

\bib     Songaila, A., Cowie, L. L., Hogan, C. J., \& Rugers, M.: 1994, 
Nature 368, 599

\bib     Songaila, A.: 1998, in Structure and evolution of the IGM from
QSO absorption lines, eds. P. Petitjean
\& S. Charlot (Fronti\`eres, Paris)

\bib     Sonneborn, G., et al.: 1999, ApJL, in preparation

\bib    Steidel, C.C.: 1993, in The evolution of galaxies and their
environment, eds. J. Shull, H.A.Thronson (Kluwer, Dordrecht), p263

\bib    Truran, J.W., Cameron, A.G.W.: 1971, Ap. Sp. Sci., 14, 179

\bib    Timmes, F.X., Truran, J.W., Lauroesch, J.T., York, D.G.: 1997, ApJ 476,
464

\bib     Tytler, D., Burles, S., Lu, L., Fan, X.-M., Wolfe, A., \& Savage, B.:
1999, AJ 117, 63

\bib Vallerga, J.V., Vedder, P.W., Carig, N., Welsh, B.Y.: 1993, ApJ 411, 729

\bib     Vangioni-Flam, E., \& Cass\'e, M.: 1995, ApJ 441, 471

\bib     Vidal-Madjar, A., Laurent, C., Bonnet, R.M., \& York, D.G.: 1977, 
ApJ 211, 91

\bib     Vidal-Madjar, A., Laurent, C., Bruston, P., \& Audouze, J.: 1978, 
ApJ 223, 589


 \bib    Vidal-Madjar, A., Ferlet, R., Lemoine, M.:  1998, Sp. Sc. Rev.
 84, 297

\bib     Vidal-Madjar, A., Lemoine, M., Ferlet, R., H\'ebrard, G., Koester, D., 
Audouze, J., Cass\'e,  M., Vangioni-Flam, E., \& Webb, J.: 1998, A\&A 338, 694

\bib    Viegas, S.M., Fria\c{c}a, A.C.S.: 1995, MNRAS 272, L35


\bib     Watson, W. D.: 1973, ApJ 182, L73

\bib     Webb, J.K., Carswell, R. F., Irwin, M. J., Penston, M. V.: 1991, 
MNRAS 250, 657

\bib     Webb, J. K., Carswell, R. F., Lanzetta, K. M., Ferlet, R., Lemoine, M., 
Vidal-Madjar, A., \& Bowen, D. V.: 1997, Nature 388, 250

\bib    Yanny, B., York, D.G.: 1992, ApJ 391, 569

\bib    York, D.G., Rogerson, J.B. Jr: 1976, ApJ 203, 378

\bib     York, D.G.: 1983, ApJ 264, 17

\bib     York, D. G., Spitzer, L., Bohlin, R. C., Hill, J., Jenkins, E. B., 
Savage, B. D., Snow, T. P.: 1983, ApJ 266, L55

\bib     York, D. G., Ratcliff, S., Blades, J. C., Wu, C. C., Cowie, L. L., 
\& Morton, D. C.: 1984, ApJ 276, 92

\bib    York, D.G., Dopita, M., Green, R., Bechtold, J.: 1986, ApJ 311, 610

\end{document}